\documentclass[runningheads]{svmult}
\usepackage{makeidx}   
\usepackage{graphicx}  
\usepackage{subeqnar}  
\usepackage{multicol}  
\usepackage{cropmark} 
\usepackage{physprbb}  
\begin{document}
\title*{Death of Stellar Baryonic Dark Matter Candidates}
\author{Katherine Freese\inst{1}
\and Brian Fields\inst{2}
\and David Graff\inst{3}}
\authorrunning{Katherine Freese et al.}
%
%
\institute{Physics Dept, University of Michigan, Ann Arbor, MI 48109, USA
\and Dept. of Astronomy, University of Illinois, Champagne-Urbana, IL, USA
\and Dept. of Astronomy, Ohio State University, Columbus, OH, USA}

\maketitle              

\def\question#1{{{\marginpar{\tiny \sc #1}}}}
\def\msun{M_\odot}
\def\zsun{Z_\odot}
\def\lsun{L_\odot}
\def\fun#1#2{\lower3.6pt\vbox{\baselineskip0pt\lineskip.9pt
  \ialign{$\mathsurround=0pt#1\hfil##\hfil$\crcr#2\crcr\sim\crcr}}}
\def\la{\mathrel{\mathpalette\fun <}}
\def\ga{\mathrel{\mathpalette\fun >}}
\def\eg{{\it e.g., }}
\def\etal{{\it et al. }}
\def\etalc{{\it et al., }}
\def\pc {{\rm pc}}
\def\mpc{{\rm Mpc}}
\def\kpc{{\rm kpc}}
\def\Mpc{{\rm Mpc}}

\def\he#1{\hbox{${}^{#1}{\rm He}$}}
\def\li#1{\hbox{${}^{#1}{\rm Li}$}}

\def\imfm{\xi_{\star}}
\def\mrem{m_{\rm rem}}
\def\avg#1{\langle #1 \rangle}
\def\sigbar{\avg{\sigma}}
\def\rbar{\hbox{$\avg{r}$}}
\def\omegam{\Omega_{\rm Macho}}
\def\omegab{\Omega_{\rm B}}
\def\omegalya{\Omega_{{\rm Ly}\alpha}}

\def\lya{Ly$\alpha$}

\def\pcite#1{(\cite{#1})}
\def\pref#1{(\ref{#1})}

\def\macho{{\sc macho}}
\def\newpage{\vfill\eject}
\def\vs{\vskip 0.2truein}
\def\gnu{\Gamma_\nu}
\def\fnu {{\cal F_\nu}}
\def\mass{m}
\def\lum{{\cal L}}
\def\imf{\Psi(\mass)}
\def\ilf{\Phi(M)}
\def\msun{M_\odot}
\def\zsun{Z_\odot}
\def\met{[M/H]}
\def\vi{(V-I)}
\def\mtot{M_{\rm tot}}
\def\mhalo{M_{\rm halo}}
\def\pp{\parshape 2 0.0truecm 16.25truecm 2truecm 14.25truecm}
\def\la{\mathrel{\mathpalette\fun <}}
\def\ga{\mathrel{\mathpalette\fun >}}
\def\fun#1#2{\lower3.6pt\vbox{\baselineskip0pt\lineskip.9pt
  \ialign{$\mathsurround=0pt#1\hfil##\hfil$\crcr#2\crcr\sim\crcr}}}
\def\ie{{\it i.e., }}
\def\eg{{\it e.g., }}
\def\etal{{\it et al. }}
\def\etalc{{\it et al., }}
\def\kpc{{\rm kpc}}
 \def\Mpc{{\rm Mpc}}
\def\mh{\mass_{\rm H}}
\def\mmax{\mass_{\rm u}}
\def\ml{\mass_{\rm l}}
\def\bc{f_{\rm cmpct}}
\def\br{f_{\rm rd}}
\def\kmsec{{\rm km/sec}}
\def\ibl{{\cal I}(b,l)}
\def\dmax{d_{\rm max}}

\begin{abstract}

The nature of the dark matter in the Universe is one of the
outstanding questions in astrophysics.  In this talk, I address
possible stellar baryonic contributions to the 50-90\%
of our Galaxy that is made of unknown dark matter.   First I show
that faint stars and brown dwarfs
constitute only a few percent of the mass of the Galaxy.
Next, I show that stellar remnants, including white dwarfs and neutron stars, 
are also insufficient in abundance to explain all the dark
matter of the Galaxy.  High energy gamma-rays observed in
HEGRA data place the most robust constraints, $\Omega_{WD} < 3 \times
10^{-3} h^{-1}$, where $h$ is the Hubble constant in units of 100 km
s$^{-1}$ Mpc$^{-1}$.  Overproduction of chemical abundances
(carbon, nitrogen, and helium) provide the most
stringent constraints, $\Omega_{WD} < 2 \times
10^{-4} h^{-1}$.  Comparison with recent updates of microlensing
data are also made.  According to the gamma-ray limit, all
Massive Compact Halo Objects seen by the experiments
(Machos) can be white dwarfs if one takes the extreme numbers;
however, from chemical overproduction limits, NOT all Machos
can be white dwarfs.  Comments on recent observations of the infrared
background and of white dwarfs are also made.
In conclusion, a nonbaryonic component in the
Halo seems to be required.

\end{abstract}

\def\question#1{{{\marginpar{\tiny \sc #1}}}}
\def\eqr#1{{Eq.\ (\ref{#1})}}
\def\be{\begin{equation}}
\def\ee{\end{equation}}

\section{Introduction}
My basic conclusions of this talk are the following:\hfil\break
I. It is looking very likely that 50-90\% of our Galaxy is made
of nonbaryonic dark matter. \hfil\break
II. Regarding the 13-17 microlensing events interpreted as being
in our Halo: these are not yet understood.

The nature of the dark matter in the universe and in our Galaxy
is one of the great unanswered questions in astrophysics.
It is clear from rotation curves of galaxies including our own
that most of the matter in galaxies is not in the form of bright
stars and instead consists of an unknown component of dark matter.
Ten years ago there were two different camps of people on this
subject.  In the first camp, there were those 
who believed that the simplest solution
would be baryonic dark matter.  In particular, the most likely
solution appeared to be stellar or substellar objects including
faint stars, brown dwarfs, white dwarfs, or neutron stars.
In the second camp, there were physicists (particularly motivated
by particle physics) who believed there must be a dominant
nonbaryonic contribution due to particles such as massive neutrinos, axions,
or supersymmetric particles. The main point of my talk is to show
that the objects preferred by the
first camp, namely stellar baryonic dark matter candidates, 
are ruled out (see also my conference proceedings in [\cite{confproc1},
\cite{confproc2}, \cite{confproc3}] for a longer
discussion).  Thus nonbaryonic dark matter seems to be favored
as explaining the mass of our Galaxy.

In 1986 Hegyi and Olive \cite{hegyi} ruled out many obvious candidates
for baryons in the Halo of our Galaxy.  They ruled out diffuse
hot gas, cool neutral hydrogen, small lumps or snowballs of
hydrogen, and rocks or dust.  In the past decade,
microlensing experiments including
MACHO, EROS, and OGLE were designed to look for MACHOs, or
Massive Compact Halo Objects, which are objects (probably baryonic)
in the $(10^{-7} -1) M_\odot$ mass range.  Instead of resolving
the dark matter puzzle, these experiments have raised new issues.
The most recent results from the MACHO experiment are discussed
by E. Aubourg in this volume.  These experiments 
(\cite{macho:2yr}, \cite{macho:6yr}, \cite{ansari}, \cite{laserre})
have the very strong result that they have
ruled out a significant component in the Halo of objects
in the mass range $(10^{-7} - 10^{-2}) M_\odot$.
These experiments have also found tens of objects not yet understood,
with a best fit mass $\sim 0.5 M_\odot$, near the mass of a white dwarf.
As a consequence, there has been a great deal of 
recent focus on a possible white dwarf component in the Halo.

In this talk,
I will discuss work showing that stellar baryonic candidates
for the dark matter are ruled out.  The stellar candidates are:
\hfil\break
1. Faint Stars. These are objects heavier than about $0.1 M_\odot$
that shine due to hydrogen burning in their cores. \hfil\break
2. Brown Dwarfs.  These are objects lighter than about $0.1 M_\odot$
that do not have hydrogen burning in their cores; hence the
easiest way to find these objects is to look for them
gravitationally (such as with microlensing
experiments).\hfil\break
3. White Dwarfs.  These are objects with mass $\sim 0.6 M_\odot$
and are the remnants of $(1-8)M_\odot$ stars. As mentioned 
above, these are the best fit to the MACHOs found by the microlensing
experiments.\hfil\break
4. Neutron Stars. These are the 1.4$M_\odot$ 
remnants of stars heavier than $8 M_\odot$.\hfil\break

As I will show in this talk,
recent work has shown that 
none of these four candidates are found in sufficient abundance
to explain the mass of our Galactic Halo.

Five years ago, many astronomers believed that the numbers
of faint stars and brown dwarfs in the Halo could be quite
substantial; in fact these appeared to be the most plausible
candidates for the Halo dark matter.  It appeared that, as one
looked at lower and lower masses, the number of stars seemed
to increase such that there could be very many low mass stars
and substellar objects.  Instead,
my work of the last few years with David Graff 
as well as the work of other authors has shown that
the first two candidates, faint stars and brown dwarfs, add 
up to less than 3\% of the mass density of the Galactic Halo.
Hubble Space Telescope data found faint stars, and one 
(\cite{bfgk}, \cite{gf96a}, \cite{gf96b}, \cite{mcs}, \cite{fbg}) can
use these data to constrain the mass density of faint stars
in the Halo. We showed that
faint stars are seen to comprise roughly 1\% of the
Halo.  Brown dwarfs are constrained by both the microlensing
experiments (as discussed above) and by our work using
a combination of parallax data and theory.  We (\cite{gf96b})
looked at the faint stars in the parallax data of \cite{dahn} and used
theory to extrapolate down into the brown dwarf regime.  We found that
brown dwarfs account for at most a few percent of the Halo.
I don't have time for further discussion of the constraints on
these two classes of candidates. I recommend the reader to 
my previous conference proceeding (\cite{confproc3}) for a fuller discussion.

\section{White Dwarfs}
Next I will proceed to discussion of white dwarfs as dark matter 
candidates. These are the most interesting stellar candidates
currently as they have the masses best fit to the MACHOs seen
by microlensing data (if one assumes that these MACHOs are indeed
in the Galactic Halo).  Is the dark matter made of white dwarfs?
There are four problems and issues that need to be addressed:
\hfil\break
1. Infrared Radiation \hfil\break
2. Initial Mass Function \hfil\break
3. Baryonic Mass Budget \hfil\break
4. Element Abundances (C, N, He$^4$)\hfil\break

We will see that, for each of these topics,
none of the expected signatures of a significant
Halo white dwarf population is found.

\subsection{Constraints from multi-TeV $\gamma$-rays seen by HEGRA}

The mere existence of multi-TeV $\gamma$-rays seen in
the HEGRA experiment places a powerful constraint on the
allowed abundance of white dwarfs.  This arises because
the progenitors of the white dwarfs would produce infrared
radiation that would prevent the $\gamma$-rays from getting here.
The $\gamma$-rays and infrared photons would interact via
$\gamma \gamma \rightarrow e^+ e^-$.  

Multi-TeV $\gamma$-rays from the blazar Mkn 501 at a redshift z=0.034
are seen in the HEGRA detector.  The
cross section for (1-10)TeV $\gamma$-rays peaks at infrared photon
energies of (0.03-3)eV.  Photons in this energy range would be
produced in abundance by the progenitor stars to white dwarfs and
neutron stars.  
By requiring that the optical depth due to
$\gamma \gamma \rightarrow e^+ e^-$ be less than one for a source at
$z=0.034$ we limit the cosmological density of stellar remnants
(Graff, Freese, Walker, and Pinsonneault \cite{gfwp}),
\begin{equation}
\label{eq:hegra}
\Omega_{\rm WD} \leq (1-3) \times 10^{-3}
h^{-1} \, .
\end{equation}
This constraint is quite robust and model independent, as it applies
to a variety of models for stellar physics, star formation rate and
redshift, mass function, and clustering.  In addition, we
can be absolutely certain the main sequence progenitors of the
white dwarfs produced light! 

Note that recent direct observations of infrared light (\cite{gorjian},
\cite{wright}) give comparable constraints on the white dwarf abundance.

\subsection{Mass Budget Issues}
First, I discuss the mass budget issues
(based on work by Fields, Freese, and Graff \cite{ffg})
general to all Halo Machos, regardless of the type of object.

\subsubsection{Contribution of Machos to the Mass Density of the Universe:}
There is a potential problem in that too many baryons are
tied up in Machos and their progenitors.
We begin by estimating the contribution of Machos to the mass density of the
universe.
Microlensing results \cite{macho:6yr} predict that the total mass
of Machos in the Galactic Halo out to 50 kpc is
$M_{\rm Macho} = (6-16) \times 10^{10} \msun \, $
(note that the new numbers are a factor of two {\it lower} than
the previous estimates of \cite{macho:2yr} and in agreement with
\cite{ansari}).
Now one can obtain a ``Macho-to-light" ratio for the Halo by
dividing by the luminosity of the Milky Way (in the B-band),
$L_{MW} \sim (1.3-2.5) \times 10^{10} L_\odot,$
to  obtain
$(M/L)_{\rm Macho} = (2.6-13)\msun/L_\odot \, .$
>From the ESO Slice Project Redshift survey \cite{zuc},
the luminosity density of the Universe in the $B$ band is
${\cal L}_B = 1.9\times 10^{8} h \ L_\odot \ {\rm Mpc}^{-3} \, .$
If we assume that the $M/L$ which we defined for the Milky
Way is typical of the Universe as a whole,
then the universal mass density of
Machos is
\begin{equation}
\label{omega}
\Omega_{\rm Macho} \equiv \rho_{\rm Macho}/ \rho_c = (0.002-0.01) \,
h^{-1} \, 
\end{equation}
where the critical density
$\rho_c \equiv
3H_0^2/8 \pi G = 2.71 \times 10^{11} \, h^2 \, M_\odot \ \Mpc^{-3}$.

We will now proceed to compare our $\omegam$
derived in Eq.\ \pref{omega} with the baryonic density in the universe,
$\omegab$, as determined by primordial nucleosynthesis.
To conservatively allow for the full range of possibilities,
we will adopt
$\omegab= (0.005-0.026) \ h^{-2} \, .$
Thus, if the Galactic halo Macho
interpretation of the microlensing
results is correct,
Machos make up an important fraction of the baryonic matter
of the Universe.
Specifically, the central values give
\begin{equation}
\label{central}
\omegam/\omegab \sim 0.4 h \, .
\end{equation}
However, the lower limit on this fraction is
considerably less restrictive,
\begin{equation}
\label{comp}
{\omegam \over \omegab} \geq 0.1 h  \, ,
\end{equation}
where we have used the lowest $\omegam$ and the
highest $\omegab$.

\subsubsection{Mass Budget constraints from
Machos as Stellar Remnants: White Dwarfs or Neutron Stars}

In general, white dwarfs, neutron stars, or black holes all came from
significantly heavier progenitors.  Hence, the excess mass left over
from the progenitors must be added to the calculation of $\Omega_{\rm
Macho}$; the excess mass then leads to stronger constraints.
Typically we find the contribution of Macho progenitors to the mass
density of the universe to be
\begin{equation}
\label{omegaprog}
\Omega_{{\rm prog}} = 4 \Omega_{{\rm Macho}} = (0.008-0.04)h^{-1} \, .
\end{equation}
The central values of all the numbers now imply
\begin{equation}
\label{toomuch}
\Omega_{\rm prog} > \Omega_B \, ,
\end{equation}
which is obviously unacceptable.  One is driven to the lowest
values of $\Omega_{\rm Macho}$ and highest value of $\Omega_B$
to avoid this problem.

\subsection{On Carbon and Nitrogen}
\label{sect:carbon}

The overproduction of carbon and/or nitrogen
produced by white dwarf progenitors is one of the
greatest difficulties faced by a white dwarf dark matter scenario,
as first noted by Smecker and Wyse \pcite{wyse}
and Gibson and Mould \pcite{gm}.
Stellar carbon yields for zero
metallicity stars are quite uncertain.
Still, according to the Van
den Hoek \& Groenewegen (1997) yields, a star of mass
2.5$\msun$ will produce about twice the
solar enrichment of carbon.
However,  stars in our galactic halo have carbon
abundance in the range $10^{-4}-10^{-2}$ solar.
Hence the ejecta of a
large population of white dwarfs would have to be removed
from the galaxy via a galactic wind.

However, carbon abundances in intermediate redshift
\lya\ forest lines have recently been measured to be
quite low, at the
$\sim 10^{-2}$ solar level
\cite{sc}, for \lya\ systems at $z \sim 3$
with column densities $N \ge 3 \times 10^{15} \, {\rm cm}^{-2}$
(for lower column densities, the mean C/H drops to $\sim 10^{-3.5}$ solar
\cite{lsbr}.

In order to maintain carbon abundances as low as $10^{-2}$ solar, only
about $10^{-2}$ of all baryons can have passed through the
intermediate mass stars that were the predecessors of Machos
(Fields, Freese, and Graff \cite{ffg}).  Such a
fraction can barely be accommodated by our results in section 4.1 for
the remnant density predicted from our extrapolation of the Macho
group results, and would be in conflict with $\Omega_{{\rm prog}}$ in
the case of a single burst of star formation.  Note that
stars heavier than 4$\msun$ may replace the carbon overproduction
problem with nitrogen overproduction 
\cite{vdhg}
\cite{lattbooth}).

Using the yields described above, we calculated the C and N that would
result from the stellar processing for a variety of initial mass
functions for the white dwarf progenitors.  We used a chemical
evolution model based on a code described in Fields \& Olive
\cite{fo98} to obtain our numerical results.  Our results are
presented in the figure.  

\begin{figure}
\centering
\includegraphics[width=.8\textwidth]{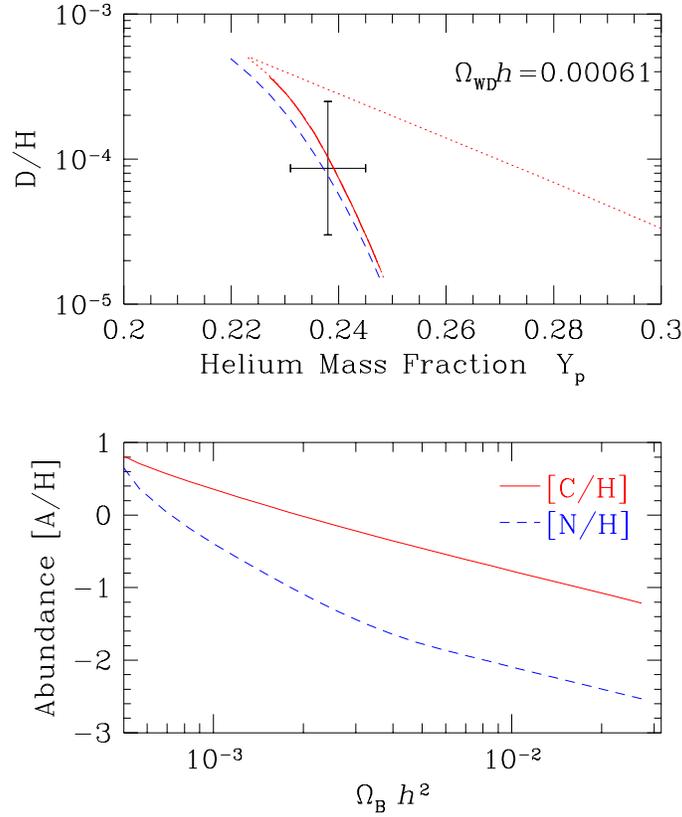}
\caption{(taken from Fields, Freese, and Graff 1999): {\bf
(a)} The D/H abundances and helium mass fraction $Y$ for models with
$\Omega_{\rm WD} h = 6.1 \times 10^{-4}$, $h=0.7$, and IMF peaked at
$2\msun$.  The red curves show the changes in primordial D and He and a
result of white dwarf production. The solid red curve is for the full
chemical evolution model, the dotted red curve is for instantaneous
recycling, and the long-dashed red curve for the burst model.  The
short-dashed blue curve shows the initial abundances; the error bars
show the range of D and He measurements.  This is the absolute minimum
$\Omega_{\rm WD}$ compatible with cosmic extrapolation of white dwarf
Machos if Machos are contained only in spiral galaxies with luminosities
similar to the Milky Way.  \hfill\break {\bf (b)} CNO abundances
produced in the same model as {\bf a}, here plotted as a function of
$\Omega_B$.  The CN abundances are
presented relative to solar via the usual notation of the form
$[{\rm C/H}]= \log_{10} \frac{{\rm C/H}}{({\rm C/H})_\odot} \, .$
The C and N production in particular are greater than 1/10
solar.}
\end{figure}

In the figure, we make the parameter choices that are in agreement with D
and He$^4$ measurements (see the discussion below) and are the least
restrictive when comparing with the Ly$\alpha$ measurements.
We take an initial mass function (IMF)
sharply peaked at 2$\msun$, so that there are very few progenitor
stars heavier than 3$\msun$ (this IMF is required by D and He$^4$
measurements).  In addition (see the figures in Fields,
Freese, and Graff \cite{ffg2}) we have considered a variety of other
parameter choices.  By comparing with the observations, we obtain the limit,
\begin{equation}
\label{eq:carbon}
\Omega_{\rm WD} h \leq 2 \times 10^{-4} \, .
\end{equation}

As a caveat, note that it is possible that carbon never leaves the
(zero metallicity)
white dwarf progenitors, so that carbon overproduction is not a
problem \cite{chabriernew}.  

\subsection{Deuterium and Helium}

Because of the uncertainty in the C and N yields from low-metallicity
stars, we have also calculated the D and He$^4$ abundances that would
be produced by white dwarf progenitors.  These are far less uncertain
as they are produced farther out from the center of the star and do
not have to be dredged up from the core.  
Panel a) in the figure displays our results.  Also shown are the initial
values from big bang nucleosynthesis and the (very generous) range of
primordial values of D and He$^4$ from observations.
>From D and He alone, we can see that the white dwarf progenitor IMF
must be peaked at low masses, $\sim 2\msun$.
We obtain
\begin{equation}
\Omega_{\rm WD} \leq 0.003 \, .
\end{equation}

\subsection{Is Dark Matter Made of White Dwarfs?}

To reiterate, there are four major problems with a white dwarf Halo:
1) infrared radiation, 2) initial mass function, 3) baryonic
mass budget, and 4) element abundances (C, N, He$^4$).  We have found
that of the expected signature of a white dwarf
population is found.  Hence white dwarfs cannot explain the full
dark matter of the Halo.

A second question remains: can the Macho data be explained by white dwarfs?
If one compares $\omegam$ with the HEGRA limit presented above
in eq.(\ref{eq:hegra}),
all MACHOs can still be white dwarfs if one takes the extreme numbers.
However, according to the limit in eq.(\ref{eq:carbon}),
NOT all Machos can be made of white dwarfs, even with the most
extreme numbers.  Gates and Gyuk \pcite{gates}
have proposed the following explanation
of the Macho data: white dwarfs in an enlarged protodisk can
explain the Macho events with $M=7 \times 10^{10} \msun$
and $\omegam \sim 3 \times 10{-3}$. These white dwarfs would
comprise (3-4)\% of the Halo density.  Even such a small
Halo fraction is hard to reconcile with eq.(\ref{eq:carbon}) above.

Recent observations \pcite{ibata1} \pcite{ibata2}
have found evidence of direct optical detections of objects that may
be Halo white dwarfs.  The situation regarding these objects
is unclear.  The objects found in the Hubble Deep
Field \pcite{ibata1} are in conflict with what was found earlier in the Luyten
survey; if the new observations are correct, large numbers of these
objects should have been found in the Luyten survey \pcite{flynn}.
Regarding the two white dwarfs found in \pcite{ibata2}, if one
takes into account the Poisson statistics, these two white dwarf
are consistent with white dwarfs from known stellar populations
\pcite{graff3}.

In any case the bulk of the dark matter has not yet been found.

\section{Zero Macho Halo?}

The possibility exists that the microlensing events that have been
interpreted as being in the Halo of the Galaxy are in fact due to some
other lensing population.  One of the most difficult aspects of
microlensing is the degeneracy of the interpretation of the data, so
that it is currently impossible to determine whether the lenses lie in
the Galactic Halo, or in the Disk of the Milky Way, or in the LMC.  In
particular, it is possible that the LMC is thicker than previously
thought so that the observed events are due to self-lensing of the
LMC.  All these possibilities are being investigated.  More data
are required in order to identify where the
lenses are.

\section{Conclusions}

Microlensing experiments have ruled out a large class of
possible baryonic dark matter components.  Microlensing experiments
have ruled out objects
in the mass range $10^{-7}\msun$ all the way up to $10^{-2}\msun$.
In this talk
I discussed the heavier possibilities in the range $10^{-2}\msun$
to a few $\msun$.  Brown dwarfs and faint stars
are ruled out as significant dark matter components; they contribute
no more than 1\% of the Halo mass density.  Stellar remnants
are not able to explain the dark matter of the Galaxy either;
none of the expected
signatures of stellar remnants, i.e., infrared radiation,
large baryonic mass budget, and C,N, and He$^4$ abundances,
are found observationally.

Hence, in conclusion, \hfill\break 1) Nonbaryonic dark matter in our
Galaxy seems to be required, and \hfill\break 2) The nature of the
Machos seen in microlensing experiments and interpreted as the dark
matter in the Halo of our Galaxy remains a mystery.  Are we driven to
primordial black holes \cite{carr} \cite{jedam},
nonbaryonic Machos (Machismos?), mirror matter Machos (\cite{mohap})
or perhaps a no-Macho Halo?


\bigskip

\section{References}

\begin{thebibliography}{99}

\bibitem{confproc1} K. Freese, B. Fields, D. Graff:
``Limits on Stellar Objects as the Dark Matter of our Galaxy:
Nonbaryonic Dark Matter seems to be Required," Proceedings of the
Nineteenth Texas Symposium on Relativistic Astrophysics,
eds. E. Aubourg, T. Montmerle, J. Paul, and P. Peter,
Nucl. Phys. B (Proc. Suppl.) 80 (2000), contr. 03/05,
astro-ph/9904401 (2000)

\bibitem{confproc2} K. Freese, B. Fields, D. Graff:
``What are Machos? Limits on Stellar Objects as the Dark Matter
of our Halo," Proceedings of the International Workshop on Aspects
of Dark Matter in Astro-Particle Physics (DARK '98), 
eds. H.V. Klapdor-Kleingrothaus and
L Baudis (World Scientific), astro-ph/9901178 (1999)

\bibitem{confproc3} K. Freese, D. Graff: Proceedings
of the International Workshop on Dark Matter in Astro- and
Particle Physics (DARK '96), eds. H.V. Klapdor-Kleingrothaus and
Y. Ramachers (World Scientific) (1997)

\bibitem{hegyi} D. Hegyi, K. Olive: Proceedings of Inner
Space/Outer Space I, eds. Kolb, Turner, Lindley, Olive, and Seckel,
(Chicago: University of Chicago Press) p. 112 (1986)

\bibitem{macho:2yr} C. Alcock, et al.: \  ApJ, 486, 697 (1997a)

\bibitem{macho:6yr} C. Alcock, et al.: \  astro-ph/0001272 (2000)

\bibitem{ansari} R. Ansari, et al.: \  A\&A, 314, 94 (1996)

\bibitem{laserre} T. Laserre: \  astro-ph/9909505 (1999)

\bibitem{bfgk} J.N. Bahcall, C. Flynn, A. Gould, 
S. Kirkahos: ApJ, 435, L51 (1994)

\bibitem{gf96a} D.S. Graff, K. Freese: ApJ, 456, L49 (1996a)

\bibitem{gf96b} D.S. Graff, K. Freese:  ApJ, 467, L65 (1996b)

\bibitem{mcs} D. M\'era, G. Chabrier, R. Schaeffer:
 Europhys.\ Lett., 33, 327 (1996)

\bibitem{fbg} C. Flynn, J. Bahcall, A. Gould:
ApJ, 466, L55 (1996)

\bibitem{dahn} C.C. Dahn, J. Liebert, H.C. Harris, H.H.
Guetter: in Proceedings of the ESO
workshop ``The bottom of the Main
sequence and Beyond" ed. C.G. Tinney (Springer Verlag,
Heidelberg) p. 239 (1995)

\bibitem{gfwp} D.S. Graff, K. Freese, T.P. Walker, M.H. Pinsonneault: 
in press, ApJ Lett, astro-ph/9903181 (1999)

\bibitem{gorjian} V. Gorjian, E.L. Wright, R.R. Chary:
astro-ph/9909428 (2000)

\bibitem{wright} E.L. Wright, E.D. Reese:  astro-ph/9912523 (1999)

\bibitem{ffg} B. Fields, K. Freese, D. Graff: New Astron, 3, 347 (1998)

\bibitem{zuc} E. Zucca, et al.:\ A\&A, 326, 477 (1997)

\bibitem{wyse} T. Smecker, R. Wyse: ApJ, 372, 448 (1991)

\bibitem{gm} B.K. Gibson, J.R. Mould:  ApJ, 482, 98 (1997)

\bibitem{sc} A. Songaila, L.L. Cowie:  AJ, 112, 335 (1996)

\bibitem{lsbr} L. Lu, W.L.W. Sargent, T.A. Barlow, M. Rauch:
astro-ph/9802189 (1998)

\bibitem{ffg2} B. Fields, K. Freese, D. Graff:
in press, ApJ (1999)

\bibitem{vdhg} L.B. van den Hoek, M.A.T. Groenewegen:
 A\&AS, 123, 305 (1997)


\bibitem{lattbooth} J.C. Lattanzio, A.I. Boothroyd:
astro-ph/9705186 (1997)

\bibitem{fo98} B.D. Fields, K.A. Olive: ApJ, 506, 177 (1998)

\bibitem{chabriernew} G. Chabrier: ApJ Lett, in press;
astro-ph/9901145 (1999)

\bibitem{gates} E. Gates, G. Gyuk: astro-ph/0004399 (2000)

\bibitem{ibata1} R. Ibata, H. Richer, R. 
Gilliland, D. Scott: astro-ph/9908270 (1999)

\bibitem{ibata2} R. Ibata, et al.: astro-ph/0002138 (2000)

\bibitem{flynn} C. Flynn,  et al.: astro-ph/9912264 (1999)

\bibitem{graff3} D. Graff: astro-ph/0005521 (2000)

\bibitem{carr} B. Carr: ARAA, 32, 531 (1994)

\bibitem{jedam} K. Jedamzik: Phys. Rev. D, 55, 5871 (1997)

\bibitem{mohap} R.N. Mohapatra, V.L. Teplitz:
astro-ph/9902085 (1999)


\end{thebibliography}
\end{document}